# A combined method for synthesis of superconducting Cu doped $Bi_2Se_3$

Meng Wang[1], Yanru Song[2], Lixing You[1], Zhuojun Li[1], Bo Gao[1], Xiaoming Xie[1], Mianheng Jiang[2]

[1]Shanghai Institute of Microsystem and Information Technology, Chinese Academy of Science, 865 Changning Road, Shanghai, China.

[2]School of Physical Science and Technology, ShanghaiTech University, 319 Yueyang Road, Shanghai, China.

## Abstract

We present a two-step technique for the synthesis of superconducting $Cu_xBi_2Se_3$. $Cu_{0.15}Bi_2Se_3$ single crystals were synthesized using the melt-growth method. Although these samples are non-superconducting, they can be employed to generate high quality superconducting samples if used as precursors in the following electrochemical synthesis step. Samples made from $Cu_{0.15}Bi_2Se_3$ reliably exhibit zero-resistance even under the non-optimal quenching condition, while samples made from pristine $Bi_2Se_3$ require fine tuning of the quenching conditions to achieve similar performance. Moreover, under the optimal quenching condition, the average superconducting shielding fraction was still lower in the samples made from pristine $Bi_2Se_3$ than in the samples made from $Cu_{0.15}Bi_2Se_3$. These results suggest that the pre-doped Cu atoms facilitate the formation of a superconducting percolation network. We also discuss the useful clues that we gathered about the locations of Cu dopants that are responsible for superconductivity.

## Introduction

Topological superconductors can host Majorana type exotic quasi-particles on its boundary. Among the various experimental efforts, the doping of topological insulators (TIs) is one of the promising ways to find materials that accommodate topological superconductivity. It has been confirmed that doping Cu or Sr into $Bi_2Se_3$ and Pd into $Bi_2Te_3$ can induce superconductivity[1-5]. $Cu_xBi_2Se_3$ has drawn the most attentions among these materials not only because it is the first superconducting material created by doping a TI, but also owing to the discovery of some exotic electronic/magnetic behaviors that may be possibly related to Majorana zero modes, such as a zero bias conductance anomaly observed in the soft point contact spectrum[6-8]. Until now, it has been possible to make superconducting $Cu_xBi_2Se_3$ samples using either the melt-growth method or the electro-chemical method[1,9,10]. The former can occasionally generate samples with high superconducting shielding fraction, but the majority of the products are only weakly superconducting, with non-vanishing resistance and very low superconducting shielding fractions. The electro-chemical method can reproducibly synthesize zero-resistance samples, but the fine tuning of annealing/quenching conditions is still necessary. To match the needs of future study, high quality $Cu_xBi_2Se_3$ samples are greatly desired. Moreover, there is still debate about the origin of the superconductivity in $Cu_xBi_2Se_3$. Initially, it was believed that the superconductivity was induced by Cu doping into the $Bi_2Se_3$ lattice[1]. Recently, a new argument that the superconductivity may be caused by a certain unknown impurity phase has emerged[10]. The microscopic origin of the superconductivity in Cu doped $Bi_2Se_3$ is still under debate[11-13]. It will surely affect whether we can expect intrinsic or extrinsic topological superconductivity in Cu doped $Bi_2Se_3$. In this letter, we report a two-step technique to synthesize superconducting $Cu_xBi_2Se_3$ using a combination of the melt-growth and electro-chemical methods. We use the melt-grown $Cu_{0.15}Bi_2Se_3$ instead of pristine $Bi_2Se_3$ as a precursor in the electro-chemical Cu intercalation process. This technique leads to a more reliable production of superconducting $Cu_xBi_2Se_3$ with a higher average superconducting shielding fraction. We also discuss a number of clues about the microscopic origin of superconductivity in the material based on these results.

## Sample Preparation

We adopted a two-step process to synthesize superconducting Cu doped $Bi_2Se_3$. In the first step, pristine $Bi_2Se_3$ and pre-doped $Bi_2Se_3$ ($Cu_{0.15}Bi_2Se_3$) single crystals were grown from stoichiometric mixtures of high purity (99.999%) Bi, Se and Cu powders. The powders were weighed and loaded into quartz ampoules inside a glove box under argon atmosphere. The ampoules were then evacuated and sealed. The melt-growth was carried out in a vertical furnace. To synthesize pristine $Bi_2Se_3$, the ampoule was heated to 750 ℃ for 12 h, followed by a slow cooling to 650 ℃ at a rate of 3 ℃/h. The synthesis was completed by turning off the furnace and letting the ampoule cool to room temperature inside. To synthesize the pre-doped $Cu_{0.15}Bi_2Se_3$, the powder mixtures were melted at 850 ℃ and held for 12 h, followed also by a slow cooling to 620 ℃ at 3 ℃/h. The ampoule was subsequently quenched in cold water. As-grown crystals exhibited a metallic appearance on the surface and were easily cleavable due to their layered structure. After the melt-growth, the next step was the electro-chemical intercalation of Cu followed by an annealing/quenching process. The pristine $Bi_2Se_3$ and the melt-grown $Cu_{0.15}Bi_2Se_3$ single crystals were cut into rectangular pieces with a typical size of $3 \times 5 \times 0.5$ mm$^3$. The Cu intercalation was carried out in a homemade electrochemical cell similar to the one described in *Kriener et al.'s* report[9]. The current flow during the reaction was around 10μA. After Cu intercalation, an annealing/quenching process is necessary to obtain superconducting samples. We performed the annealing process in a box furnace and also in a vertical tube-furnace. The major advantage of using the vertical tube-furnace is that the sample could be dropped much faster into ice-water for quenching. The details of how quenching influenced the superconducting behavior of the samples will be discussed below.

**Results**

**X-ray diffraction and electrical resistivity analysis**

Before we describe the superconducting behavior of the samples made from pristine $Bi_2Se_3$ and from the melt-grown $Cu_{0.15}Bi_2Se_3$ through the electro-chemical intercalation process, we first present the structural and the electrical properties of the melt-grown $Cu_{0.15}Bi_2Se_3$ samples. Fig.1a depicts the powder X-ray diffraction pattern of a typical $Cu_{0.15}Bi_2Se_3$ sample, in which no obvious impurity phase can be found. This result agrees with the previous report that a pure single crystal sample can be obtained as long as the nominal Cu ratio is not larger than 0.15[1]. The inset of Fig. 1a shows the single crystal X-ray diffraction patterns for three samples: the pristine $Bi_2Se_3$ (red), the melt-grown $Cu_{0.15}Bi_2Se_3$ (blue) and the superconducting $Cu_xBi_2Se_3$ (green) made by the electro-chemical method. The pristine $Bi_2Se_3$ was found to have a c-axis lattice constant of 28.63(0) Å, while the melt-grown $Cu_{0.15}Bi_2Se_3$ (28.67(0) Å) and the superconducting $Cu_xBi_2Se_3$ (28.68(4) Å) samples exhibited clear c-axis lattice expansion. Though the melt-grown $Cu_{0.15}Bi_2Se_3$ samples manifest also c-axis lattice expansion, most of them are non-superconducting, and a few exceptions showed only a tiny resistance drop at ~3 K. Fig. 1b depicts the typical temperature dependent resistivity data (black) of the melt-grown $Cu_{0.15}Bi_2Se_3$. It exhibited only metallic behavior, and no superconducting transition could be observed down to 1.8 K. We also re-annealed/quenched these samples under the same conditions as we would employ after the electro-chemical intercalation process, no superconducting transition could be found in the majority of samples either, as shown by the curve in red.

**Annealing in a box-furnace**

To make superconducting samples, we used both the pristine $Bi_2Se_3$ and the melt-grown $Cu_{0.15}Bi_2Se_3$ as precursors in the electro-chemical Cu intercalation process. In the first set of experiments, the intercalated samples were annealed in a box-furnace and then quenched in ice-water. The ampoules were usually filled with 0.6 bar helium gas to accelerate the cooling of the samples during the quenching process. The annealing temperature for the samples made from $Cu_{0.15}Bi_2Se_3$ was varied between 530 ℃ and 560 ℃, and nearly all the obtained samples reached zero-resistance. In contrast, it was difficult to produce zero-resistance in the samples made from pristine $Bi_2Se_3$. To find the optimal annealing condition for the samples made from pristine $Bi_2Se_3$, we varied the annealing temperature from 450 ℃ to 600 ℃. No superconducting transition was observed when the samples were annealed at a temperature lower than 500 ℃. When the annealing temperature was higher than 580 ℃, the samples tended to melt during the annealing and became powder like after quenching. The shielding fraction of these samples was also quite low. When annealed between 500 ℃ and 560 ℃, the samples made from pristine $Bi_2Se_3$ became partially superconducting; occasionally we could obtain a sample that reached zero-resistance but it was hard to reproduce the result. Figure 2a and 2b depict the results of resistance and magnetic susceptibility measurements for two typical samples, one made from pristine $Bi_2Se_3$ and the other one made from $Cu_{0.15}Bi_2Se_3$. The magnetic measurements were carried out at 2 K under the zero-field-cooling condition (ZFC). The magnetic susceptibility of the samples was calculated by linear fitting of the low-field part of the M-H curves. The sample made from $Cu_{0.15}Bi_2Se_3$ exhibited a shielding fraction of ~42%, while that of the sample made from pristine $Bi_2Se_3$ was only 13%. It is clear that the synthesis of superconducting Cu doped $Bi_2Se_3$ was more reliable using the melt-grown $Cu_{0.15}Bi_2Se_3$ as precursor.

**Annealing in a vertical furnace**

Because previous reports have shown that samples made from pristine $Bi_2Se_3$ can also reach zero resistance, we continued to optimize our annealing procedure. It is known that the quenching temperature plays an important role in obtaining superconducting $Cu_xBi_2Se_3$ samples[1,9,10,14]. When using the box-furnace, we noticed that it would take more than one second before the quartz ampoule could be immersed in the ice water, and the sample temperature could drop significantly in this time before quenching actually occurred. We therefore switched to a vertical furnace for the annealing. The quartz ampoule was hung inside the furnace using a metal wire. After breaking of the metal wire, the ampoule dropped into the ice water in less than half a second. We carried out the second set of experiments using this optimized quenching process. For both pristine $Bi_2Se_3$ and the melt-grown $Cu_{0.15}Bi_2Se_3$ precursors, setting the annealing temperature to 560 ℃ produced samples that could usually reach zero resistance. However, measurements of 38 samples revealed that the samples made from $Cu_{0.15}Bi_2Se_3$ still had a higher average superconducting shield fraction. Figure 3 depicts the shielding fraction data collected from these samples. All the data were calculated from M-H curves taken at 2 K under the ZFC condition. The columns corresponding to various ranges of superconducting shielding fraction have been separated for clarity. The Gaussian fitting of the histogram clearly reveals that the samples made from the melt-grown $Cu_{0.15}Bi_2Se_3$ had a higher average shielding fraction than the samples made from pristine $Bi_2Se_3$. Moreover, all the samples having a shielding fraction larger than 40% were made from the melt-grown $Cu_{0.15}Bi_2Se_3$. We note that the difference between the average shielding fractions of the two types of samples was roughly 10%, which is much larger than the average shielding fraction measured for re-annealed $Cu_{0.15}Bi_2Se_3$ without performing the electro-chemical Cu intercalation. This suggests that the difference did not arise from residual superconducting parts of the (re-annealed) $Cu_{0.15}Bi_2Se_3$.

## Discussions

Because strong c-axis lattice expansion had been previously observed in the superconducting $Cu_xBi_2Se_3$ samples[1], it was believed that the superconductivity of $Cu_xBi_2Se_3$ was caused by Cu intercalation into van der Waals (VDW) gaps. However, there is no solid proof supporting this argument. On the contrary, several works have shown that Cu intercalation does not necessarily bring superconductivity[9,15,16]. It is now well known that an appropriate annealing/quenching process is the key to obtain superconducting $Cu_xBi_2Se_3$ samples, regardless of whether the melt-growth or the electro-chemical method is used. It suggests that a metastable state is responsible for the superconductivity in Cu doped $Bi_2Se_3$[10]. The existence of such a metastable state was also verified in our experiments. The samples made from pristine $Bi_2Se_3$, if annealed in the vertical furnace, definitively exhibited a higher superconducting shielding fraction than those annealed in the box-furnace. The quenching speed and the actual quenching temperature decided how well the superconducting metastable state could be preserved. Currently, there is still debate about the nature of such a metastable state. One assumption is that the superconductivity comes from regions with high Cu concentration that have the same phase as pristine $Bi_2Se_3$, and upon cooling the Cu atoms in these regions may take other doping positions that are energetically more favorable but cannot induce superconductivity. There are also other assumptions, such as that the superconductivity comes from an unknown impurity phase, similar to the situation in $K_xFe_{2-y}Se_2$[17]. The exact answer to this question is still pending.

Despite the lack of detailed information about the locations of Cu dopants and their roles in inducing superconductivity, we get a number of clues to these questions based on our experimental findings. The fact that the samples made from the melt-grown $Cu_{0.15}Bi_2Se_3$ reliably reached zero-resistance even under the non-optimal quenching condition suggests that copper atoms in the pre-doped $Cu_{0.15}Bi_2Se_3$ facilitate the formation of a superconducting percolation network. Currently we have two assumptions about the locations of these copper atoms that eventually contribute to the superconductivity. One is that these copper atoms exist in an unknown impurity phase; the other one is that they stay in some metastable positions in the $Bi_2Se_3$ lattice. We think that the first assumption is less likely because no impurity peaks had been observed from the powder X-ray diffraction patterns, and also the pre-doped $Cu_{0.15}Bi_2Se_3$ are non-superconducting. Up to now no direct observation of such superconducting impurity phase has ever been reported. Concerning the second assumption, though various locations of Cu dopants have been experimentally verified[16,18,19], including the interstitial sites and the intercalation sites in van der Waals gap, it is not clear yet which location is responsible for the superconductivity. Due to the large separation between adjacent quintuple layers, metal atoms may diffuse quite freely inside VDW gaps[18,20,21]. Wang et al. calculated the activation energy $E_{//}$ for Cu atom diffusion in the direction parallel to the surface[18]. It was found that $E_{//}$ was only 317 meV, from which the in-plain Cu atom diffusion coefficient $D_{//}$ at the annealing temperature 560 ℃ was calculated to be in the order of $10^{-4}$ $cm^2$/sec. Using $<r^2> = 4Dt$, a typical annealing time t around one hour leads to a mean diffusion distance r in the order of one centimeter, which is larger than the size of our samples. It means that the annealing will likely generate a homogenous distribution of Cu atoms inside VDW gaps. In this case, if we assume that the metastable positions that are responsible for superconductivity are located inside VDW gaps, we feel that it does not provide a natural explanation to why the samples made from the melt-grown $Cu_{0.15}Bi_2Se_3$ can always reach zero-resistance when annealed in the box-furnace. Since the annealing process tends to generate a homogenous distribution of Cu atoms inside VDW gaps, if the samples made from the melt-grown $Cu_{0.15}Bi_2Se_3$ can always reach zero-resistance, we expect that a reasonable amount of samples made from pristine $Bi_2Se_3$ would also reach zero-resistance when annealed in the box-furnace. This is in contradictory with our observations. In addition, Wang et al. has found that the energy of Cu dopants at the intercalation sites is lower than the interstitial sites[18], which makes the assumption that Cu intercalation brings superconductivity fit less well with the picture of a metastable state that usually refers to a state with higher energy. The possibility that the metastable positions of copper atoms are located outside VDW gaps therefore deserves consideration. Cu atoms in these positions are less mobile than the positions inside VDW gaps. Assuming that there is an energy barrier between the intercalation positions and the metastable positions, the annealing will drive some Cu dopants to overcome the energy barrier and displace to the localized metastable positions. The existence of pre-doped Cu atoms in these positions means that more local regions in the sample can match the minimum Cu abundance requirements for realizing superconductivity. It is therefore easy to form a superconducting percolation network in the samples made from the melt-grown $Cu_{0.15}Bi_2Se_3$. Of course, we cannot completely exclude the possibility that the metastable positions for Cu atoms also exist inside van der Waals gaps since the lattice defects such as Se vacancies can reduce the diffusion coefficient of Cu atoms in VDW gaps, making Cu atoms less mobile in VDW gaps. Experiments using advanced structure characterization tools are in progress to further explore the nature of the metastable positions.

To conclude, we have found a combined method using both the melt-growth and the electro-chemical Cu intercalation to synthesize superconducting Cu doped $Bi_2Se_3$ in a reliable manner. The samples made from the pre-doped $Cu_{0.15}Bi_2Se_3$ exhibit higher tolerance to the quenching conditions. It suggests that the pre-doped Cu atoms facilitate the formation of a superconducting percolation network. Results under different quenching conditions prove that the superconductivity in Cu doped $Bi_2Se_3$ originates from a metastable state. Though the exact location of the metastable state is not yet clear, the possibility that the metastable positions are outside VDW gaps deserves consideration.

# Acknowledgement

We thank Shan Qiao, Mao Ye, He Tian and Wei Li for helpful discussions. We acknowledge the support from the "Strategic Priority Research Program (B)" of the Chinese Academy of Sciences under Grant No. XDB04010600; from the National Natural Science Foundation of China under Grant No. 11204337 and 11374321; and from Helmholtz Association through the Virtual Institute for Topological Insulators (VITI).

# Author Information

## Affiliations

**Shanghai Institute of Microsystem and Information Technology, Chinese Academy of Science, 865 Changning Road, Shanghai, China.**

Meng Wang, Lixing You, Zhoujun Li, Bo GAO, Xiaoming Xie

**School of Physical Science and Technology, ShanghaiTech University, 319 Yueyang Road, Shanghai, China.**

Yanru Song, Mianheng Jiang

## Contributions

M.Wang and Z. Li performed the experiments. Y. Song helped to make the figures. L. You, X. Xie and M. Jiang contributed to the discussions. Z. Li and B. Gao wrote the manuscript. B. GAO refined the final manuscript.

## Competing interests

The authors declare no competing financial interests.

## Corresponding Authors

Correspondence to [Zhoujun Li](#) and [Bo GAO](#)

**Figure captions:**

**Figure 1**: (a) Powder X-ray diffraction pattern of a melt-grown $Cu_{0.15}Bi_2Se_3$ sample, no impurity phase is visible. Inset: single crystal X-ray diffraction patterns of pristine $Bi_2Se_3$ (red), the melt-grown $Cu_{0.15}Bi_2Se_3$ (blue) and the superconducting $Cu_xBi_2Se_3$ made through electro-chemical Cu intercalation (green). C-axis lattice expansion is obvious for the last two samples. (b) Temperature dependent resistivity curves for a melt-grown $Cu_{0.15}Bi_2Se_3$ sample measured before (black) and after (red) the re-annealing and quenching. Inset: Zoom-in of the curves near the $Cu_xBi_2Se_3$ superconducting transition temperature.

**Figure 2**: (a) Comparison of the temperature dependent resistivity curves of two superconducting samples made from pristine $Bi_2Se_3$ (blue) and from the melt-grown $Cu_{0.15}Bi_2Se_3$ (red), the annealing for both samples was carried out in a box-furnace. (b) Magnetizations curves of the same two samples measured at 2K for H//ab under zero-field-cooling condition. The dashed lines are the linear fitting of the low field part of the M-H curves.

**Figure 3**: Histogram of the shielding fractions for samples made from pristine $Bi_2Se_3$ and from the melt-grown $Cu_{0.15}Bi_2Se_3$. The columns are separated for clarity. Red and Blue solid lines are the Gaussian fitting to the histogram.

Fig. 1

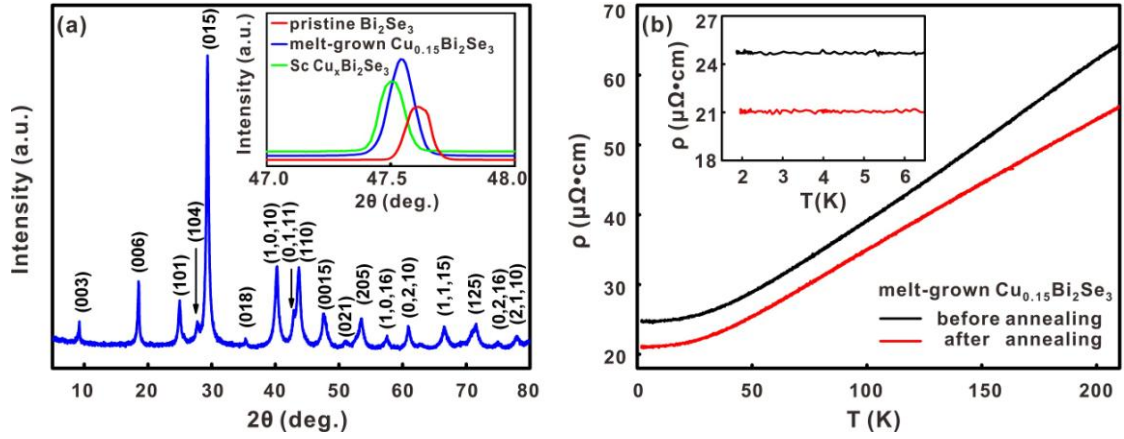

Fig. 2

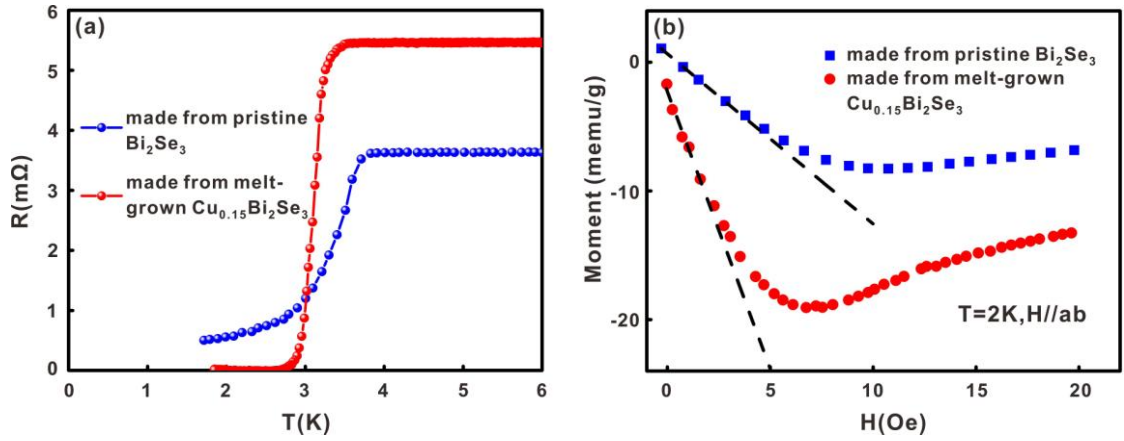

Fig. 3

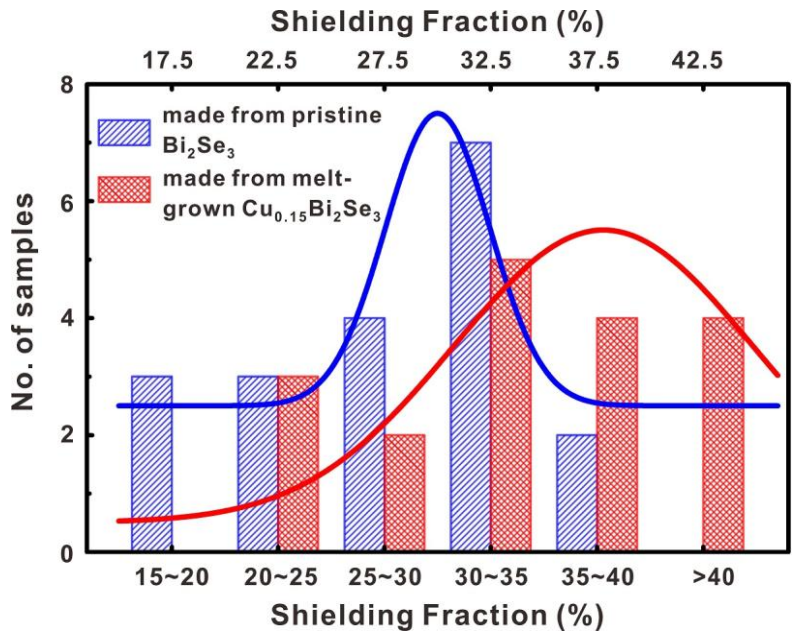